\documentclass[aps,twocolumn,superscriptaddress,notitlepage,longbibliography]{revtex4-1}
\usepackage{amsmath,amssymb,bm,bbm,mathrsfs,braket,color,graphicx,float,comment,amsfonts,dsfont}
\usepackage[colorlinks,citecolor=blue,urlcolor=blue]{hyperref}
\usepackage{multirow}

\DeclareMathAlphabet\mathbfcal{OMS}{cmsy}{b}{n}

\usepackage{extarrows} 
\usepackage{mathrsfs}

\definecolor{purple}{rgb}{.8, 0., 0.}
\definecolor{grun}{rgb}{0.0, 0.7, 0.0}

\begin{document}

\title{Exact solutions for topological surface states of three-dimensional lattice models}

\author{Matias Mustonen}
\address{Computational Physics Laboratory, Physics Unit, Faculty of Engineering and Natural Sciences, Tampere University, FI-33014 Tampere, Finland}

\author{Teemu Ojanen}
\address{Computational Physics Laboratory, Physics Unit, Faculty of Engineering and Natural Sciences, Tampere University, FI-33014 Tampere, Finland}
\address{Helsinki Institute of Physics, University of Helsinki, Helsinki FI-00014, Finland}

\author{Ali G. Moghaddam}\email{ali.moghaddam@aalto.fi}
\address{Department of Applied Physics, Aalto University, 02150 Espoo, Finland}
\address{Computational Physics Laboratory, Physics Unit, Faculty of Engineering and Natural Sciences, Tampere University, FI-33014 Tampere, Finland}
\address{Helsinki Institute of Physics, University of Helsinki, Helsinki FI-00014, Finland}

\begin{abstract}
In this work, we employ a generalized transfer matrix method that provides exact analytical and numerical solutions for lattice versions of topological models with surface termination in one direction. We construct a generalized eigenvalue equation, equivalent to the conventional transfer matrix, which neither suffers from nor requires singular (non-invertible) inter-layer hopping matrices. This contrasts with some previous approaches that use the transfer matrix technique to obtain analytical expressions for surface states, which only focus on singular cases. We then apply this formalism to derive, with exactness, the topological surface states and Fermi arc states in two prototypical topological models: the 3D Bernevig-Hughes-Zhang model and a lattice model exhibiting Weyl semimetal behavior. Our results show that the surface states and bulk bands, across the projected 2D Brillouin zone, agree perfectly with those obtained through direct numerical diagonalization of the corresponding Hamiltonians in a slab geometry. This highlights that the generalized transfer matrix method is not only a powerful tool but also a highly efficient alternative to fully numerical methods for investigating surface physics and interfaces in topological systems, particularly when it is required to go beyond low-energy effective descriptions.
\end{abstract}

\maketitle

\section{Introduction}

One of the distinguishing features of topological phases is the presence of robust boundary states, which are gapless and localized at the boundary of the system \cite{Kane2010RMP}. Unlike the boundary modes in non-topological insulating states, these surface states persist in the presence of perturbations and disorder, as long as the bulk gap and the corresponding topological insulating phase endure \cite{bernevig2013book}. The existence and robustness of these boundary states are fundamentally linked to the nontrivial topology of the bulk system \cite{qi-zhang-2011}. This relationship embodies the principle of bulk–boundary correspondence \cite{Ryu2016}, first demonstrated in quantum Hall systems \cite{Hatsugai1993} and later extended to other classes of materials, including particle–hole symmetric systems \cite{Ryu2002} and various two-dimensional topological insulators \cite{graf2013bulk}. During the past decade, most theoretical studies have focused on deriving the dispersion relations of topological surface states using low-energy effective models, first introduced in Ref. \cite{zhang2009topological} and further developed in Ref. \cite{Zhang2010model}. These effective models have been widely employed to explore the qualitative features of the spin structure of surface states in bulk systems \cite{Brouwer2012} as well as in thin-film geometries \cite{shen2012book,shen2024}. Alternatively, most of the works often relied on full numerical methods for tight-binding models and any other realistic lattice models of topological states \cite{Teo2008,Hassan2012}.

The transfer matrix technique has recently been reintroduced as a powerful method for analyzing topological boundary modes in tight-binding systems \cite{Dwivedi2016}. Subsequent studies have extended this approach to non-Hermitian models \cite{Dwivedi2019}, adapted it for unitary systems such as Floquet models \cite{Delplace2015}, and also systems with complex band structures \cite{Reuter2017review}. These methods build on the original concept of calculating a surface's electronic structure, along with bulk states, by solving recursive equations formulated using the transfer matrix \cite{Joannopoulos1981, Schulman1982}. However, most studies to date have focused on applying these methods to two-dimensional (2D) or two-band models, such as the Qi-Wu-Zhang (QWZ) model, which serves as a toy model for the Chern insulating phase \cite{QWZ2006}. Moreover, recent research has primarily aimed to develop a formalism capable of handling cases with singular (non-invertible) inter-layer hopping matrices. Here, following the original ideas from Ref. \cite{Joannopoulos1981}, we establish a generalized version of the transfer matrix method that is well-suited for deriving analytical results for surface states.
This approach ensures that our method works effectively for both singular and non-singular forms of the inter-layer hopping matrix, neither suffering from nor relying on the presence of a singular matrix.
It is worth noting that similar approaches, designed to circumvent the singular case, have been employed in various contexts, notably in ab initio studies \cite{Fujimoto2003, Kong2006} and in efficient numerical schemes \cite{Tsukamoto2018}, particularly for investigating transport properties \cite{Khomyakov2004, Sorensen2008}.
In our work, we complement this method by incorporating a convenient form of boundary condition that accounts for the decaying nature of surface states into the bulk. This makes the method suitable for identifying surface states in a semi-infinite geometry. {{Moreover, by obtaining the transfer-matrix eigenvalues and eigenstates and applying the boundary conditions, our results can be directly used to compute relevant physical quantities, such as the contributions of surface states to conductance or their spin textures.}}

Here, instead of the standard equation in terms of the transfer matrix ${\mathbbm T}$, we construct an equivalent relation with two matrices, ${\mathbbm S}$ and ${\mathbbm R}$, such that ${\mathbbm R}{\mathbbm T}={\mathbbm S}$, with ${\mathbbm R}$ being responsible for resolving the non-invertibility issue. Then, we find the eigenvalues and eigenvectors of the generalized eigenvalue problem associated with the two matrices ${\mathbbm S}$ and ${\mathbbm R}$. The surface states, as well as the delocalized bulk states, can all be decomposed in terms of the resulting eigenstates. Therefore, by applying the boundary conditions corresponding to surface termination, we derive the necessary and sufficient conditions from which the energies of surface states, as well as the edges of bulk bands (in the projected 2D Brillouin zone of the surface), can be determined. We apply this formalism to well-known lattice models for topological insulators and semimetals, including the 3D Bernevig-Hughes-Zhang (BHZ) model and a minimal two-band model for 3D Weyl semimetals.

It is worth noting that the transfer matrix method is not only a highly efficient alternative to direct diagonalization of real-space Hamiltonians, but in the case of semi-infinite geometries or regions far from surfaces, it serves as the only exact method. Brute-force diagonalization is inherently limited to finite-size systems and can only provide approximate descriptions of semi-infinite or isolated surface states. In contrast, the transfer matrix method offers an exact formalism for studying surface physics in infinite systems, beyond the capabilities of any brute-force approach. Even when analytical evaluation of this exact method is not feasible, the formalism can still be applied numerically, maintaining its efficiency and precision.

\section{Transfer matrix formulation}\label{sec:formulation}
\begin{figure}[t!]
\includegraphics[width=0.9\columnwidth]{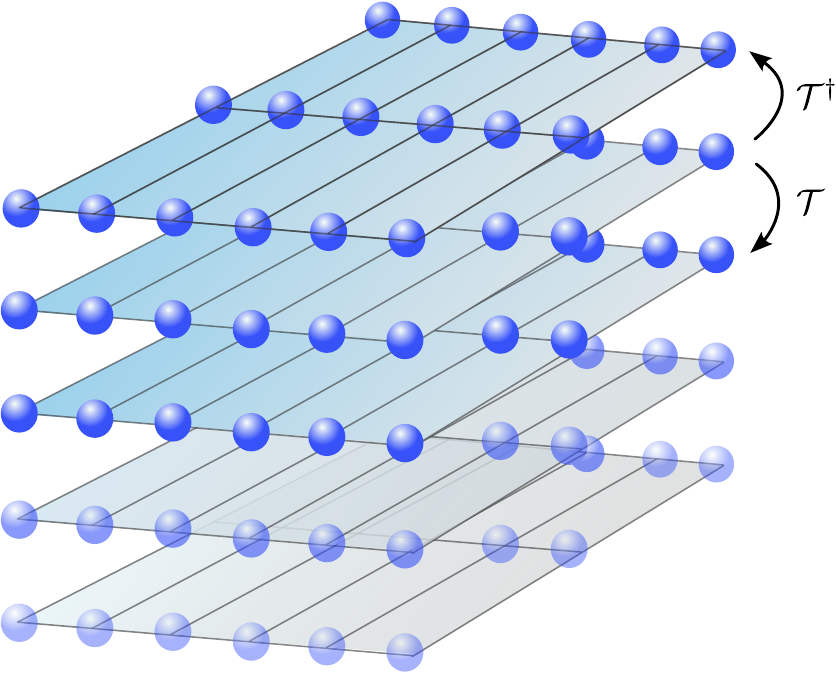}
\caption{
A schematic illustration of layered atomic structure used to construct the recursive relations of inter-layer couplings, from which the transfer matrix is constructed. ${\cal T}$ and ${\cal T}^\dag$ denote the hopping matrix between adjacent layers.  
}
\label{fig1}
\end{figure}

To utilize the transfer matrix technique, we consider two specific geometries: slabs and semi-infinite systems. In these two situations, one particular dimension, which we refer to as the vertical direction, features terminations on one or both sides. Then, assuming periodicity along all other dimensions, we will have one or two surfaces for slabs and semi-infinite solids, respectively. We further assume a layered structure along the vertical direction as depicted in Fig. \ref{fig1}, where there are only nearest-neighbor couplings between adjacent layers.

Under the above conditions, the tight-binding model can be equivalently described by a set of coupled recursive equations
\begin{equation}
    (\varepsilon-{\cal M}) \ket{\psi_{n}({\bf k}_{\parallel})}={\cal T}
    \ket{\psi_{n+1}({\bf k}_{\parallel})} +
    {\cal T}^\dag
    \ket{\psi_{n-1}({\bf k}_{\parallel})} ,
\label{eq:transer1}
\end{equation}
where matrices ${\cal M}$ and ${\cal T}$ represent intra- and inter-layer hoppings, respectively. Both of these matrices explicitly depend on the transverse momentum ${\bf k}_\parallel$ provided that we have periodicity in all transverse directions. 
For non-singular inter-layer hopping matrix ${\cal T}$, we can recast the equation above as\begin{align}
    \begin{pmatrix}
    \ket{\psi_{n+1}({\bf k}_\parallel)} \\ \ket{\psi_{n}({\bf k}_\parallel)}
    \end{pmatrix}
    = {\mathbbm T}
    \begin{pmatrix}
    \ket{\psi_{n}({\bf k}_\parallel)} \\ \ket{\psi_{n-1}({\bf k}_\parallel)}
    \end{pmatrix},\label{eq:transer2}
\end{align}
which basically defines the transfer matrix 
\begin{align}
{\mathbbm T}({\varepsilon,{\bf k}_\parallel})
=
    \begin{pmatrix}
 {\cal T}^{-1}(\varepsilon-{\cal M}) &   -{\cal T}^{-1}{\cal T}^\dag \\
 {\mathbbm 1} & {0}
    \end{pmatrix}.
    \label{eq:transfer_matrix_def}
\end{align}
For a model with $\mu$ different bands, 
the transfer matrix ${\mathbbm T}$ is a $2\mu\times 2\mu$ matrix.
The special case of non-invertible ${\cal T}$ such that ${\cal T}^2=0$ (or more generally ${\cal T}^l=0$ for some integer $l>1$.) has been extensively
explored in \cite{Dwivedi2016,Dwivedi2019}. Here we will mostly rely on the more generic 
and common cases where ${\rm det}{\cal T}\neq 0$. Nonetheless, as we will see in the following,
we can still obtain the surface states of the system even when ${\cal T}$ is not invertible.
To this end, we slightly modify the transfer matrix equation \eqref{eq:transer2} as
\begin{align}
   {\mathbbm R}
    \begin{pmatrix}
    \ket{\psi_{n+1}({\bf k}_\parallel)} \\ \ket{\psi_{n}({\bf k}_\parallel)}
    \end{pmatrix}
    = {\mathbbm S}
    \begin{pmatrix}
    \ket{\psi_{n}({\bf k}_\parallel)} \\ \ket{\psi_{n-1}({\bf k}_\parallel)}
    \end{pmatrix},\label{eq:S_matrix}
\end{align}
with \begin{align}
{\mathbbm S}
=
    \begin{pmatrix}
 \varepsilon-{\cal M} &   -{\cal T}^\dag \\
 {\mathbbm 1} & {0}
    \end{pmatrix},\quad
      {\mathbbm R} = \begin{pmatrix}
        {\cal T} &0 \\ 0& {\mathbbm 1}     
    \end{pmatrix},
    \label{eq:S_matrix_def}
\end{align} 
where both matrices ${\mathbbm S}$ and ${\mathbbm R}$ are explicit functions of energy $\varepsilon$ and transverse momentum ${\bf k}_\parallel$. One can readily check that 
Eq. \eqref{eq:S_matrix_def} is basically equivalent to Eq. \eqref{eq:transer2} yet more generic 
as it does not require ${\cal T}$ to be invertible.

Typically, by diagonalizing the transfer matrix ${\mathbbm T}$,
we can obtain the electronic structure of the system including boundary modes or surface states 
which are localized at the boundaries and exponentially vanishes by going deep into the bulk \cite{Joannopoulos1981}.
As one expects, the eigensolutions of the transfer matrix \eqref{eq:transfer_matrix_def} are the same
as the generalized eigenvalue problem corresponding to Eq. \eqref{eq:S_matrix} as
\begin{align}
    {\mathbbm S} \ket{\Phi_\alpha} = 
    \lambda_\alpha     {\mathbbm R}
     \ket{\Phi_\alpha}.
     \label{eq:generalized_eigen_S_R}
\end{align}
By plugging in the matrix structure given by Eq.~\eqref{eq:S_matrix_def}, we see that all eigenstates can be decomposed to two sub-columns as
\begin{equation}
    \ket{\Phi_\alpha} = 
    \begin{pmatrix}
        \lambda_\alpha \ket{\varphi_\alpha} \\
        \ket{\varphi_\alpha}
    \end{pmatrix}, \label{eq:Phi-decomposed}
\end{equation}
which gives rise to 
\begin{equation}
  \lambda_\alpha  [( \varepsilon-{\cal M})-{\cal T}
    \lambda_\alpha] \ket{\varphi_\alpha} - {\cal T}^\dag \ket{\varphi_\alpha} =0, \label{eq:half_kets}
\end{equation}
and a corresponding characteristic equation
\begin{align}
    {\rm det}\left[
        \lambda^2 {\cal T}-\lambda \left(\varepsilon-{\cal M}\right) + {\cal T}^\dag \right] = 0. \label{eq:main-characteristic}
\end{align}
In the next section, we will use this to obtain the surface states of some well-known tight-binding models  hosting topological phases.

Since we are dealing with a static time-independent setting, the conservation of the current density requires that the transfer matrix has a determinant equal to unity.
Furthermore, one can show that the eigenvalues of the transfer matrix
consists of pairs of complex numbers $\lambda_\alpha$ and $\lambda_{\alpha'}$
with the property ($\lambda_\alpha^\ast \lambda_{\alpha'}=1$) \cite{Joannopoulos1981}, meaning that
for any given eigenvalue $\lambda$, there must be a complementary eigenvalue $1/\lambda^\ast$.
This becomes very crucial in finding the surface states of the system
which decay from the surface into the bulk of the material. In fact, 
surface states correspond to eigenvalues whose absolute value is smaller than unity ($|\lambda_\alpha|<1$)
in semi-infinite geometry as considered here.
Due to the above condition, there are exactly equal number of eigenvalues of 
with absolute value larger than unity ($|\lambda_\alpha|>1$),
which can be interpreted as surface states of a complementary system
where the surface is located at the opposite end of the semi-infinite geometry.
The bulk states correspond to the eigenvalues with unit absolute value ($|\lambda_\alpha|=1$)
which are not localized on surfaces.
Now, quite generally, we can always divide the
diagonalizing matrix of the ${\mathbbm T}$ into two parts as
\begin{align}
    {\cal V}_{\mathbbm T} = \left(
         {\cal V}_{\leq 1},{\cal V}_{\geq 1} \right)
\end{align}
where ${\cal V}_{\leq 1}$ and ${\cal V}_{\geq 1}$ are matrices 
with dimensions $2\mu\times \mu$ whose columns are the eigenvectors of 
${\mathbbm T}$ corresponding to the eigenvalues with absolute values smaller and larger 
than unity, respectively (For those eigenvalues which $|\lambda_\alpha|=1$ the division would be quite arbitrary).
One natural choice in the above decomposition is to sort the eigenvalues in ascending order based on their absolute values,
such that
\begin{align}
    {\cal V}_{\leq 1} &= \left(\ket{\Phi_{1}},\cdots,\ket{\Phi_\mu}\right),
    \\ 
    {\cal V}_{\geq 1} &= \left(\ket{\Phi_{\mu+1}},\cdots,\ket{\Phi_{2\mu}}\right).
\end{align}
Then, the most generic solution for the surface states of the system reads
\begin{align}
    \ket{\Psi_n}_{\rm sur}=
    \begin{pmatrix}
        \ket{\psi_{n+1}} \\ \ket{\psi_{n}}
    \end{pmatrix}
= \sum_{\alpha=1}^{\mu}c_\alpha\lambda_\alpha^n \ket{\Phi_\alpha},
\label{eq:surface_state_Psi}
\end{align}
with coefficients $c_\alpha$. At this point, it is worth reminding that both the eigenvalues and eigenvectors are functions of the 
transverse momentum ${\bf k}_\parallel$ and the energy $\varepsilon$.
In order to have a complete set of
equations from which we can obtain unique solutions for the surface states and their corresponding energies,
we would also require boundary conditions. 
Assuming surface termination of the system at $n=0$, 
such that $\ket{\psi_{-1}}=0$, Eq. \eqref{eq:transer1} at the surface reduces to
\begin{equation}
    (\varepsilon-{\cal M}) \ket{\psi_{0}({\bf k}_{\parallel})}={\cal T}
    \ket{\psi_{1}({\bf k}_{\parallel})}  ,
\label{eq:transer-surface}
\end{equation}
which serves as the boundary condition for the surface states.
We can re-write Eq. \eqref{eq:transer-surface} invoking the 
surface states expression given by Eq. \eqref{eq:surface_state_Psi} as
\begin{align}
    & \left(
    -{\cal T} \,,\:
    \varepsilon-{\cal M} \right)\ket{\Psi_0}_{\rm sur}\nonumber\\
    &\qquad=
    \sum_{\alpha=1}^{\mu}  
    \left(
    -{\cal T} \,,\:
    \varepsilon-{\cal M} 
    \right)
    \ket{\Phi_\alpha} \, c_\alpha 
    \nonumber\\
    &\qquad=
     \left(
    -{\cal T} \,,\:
    \varepsilon-{\cal M} 
    \right)
    {\cal V}_{\leq 1} \ket{\{c_\mu\}}= 0,
\end{align}
with $\ket{\{c_\mu\}}$ being the column vector consisting of coefficients $c_\alpha$ ($1\leq \alpha\leq \mu$).
The above equation can have nontrivial solutions ($\ket{\{c_\mu\}}\neq0$) only if 
\begin{align}
    {\rm det}\left[ 
    \left(
    -{\cal T} \,,\:
    \varepsilon-{\cal M} 
    \right)
    {\cal V}_{\leq 1}\right] = 0,\label{eq:decay-criterion-old}
\end{align}
from which the surface state energies $\varepsilon_{\rm sur}$ can be obtained. The decaying surface states criterion given by Eq. \eqref{eq:decay-criterion-old} can be further simplified to 
\begin{equation}
 {\rm det}\left[\frac{1}{\lambda_1}  \ket{\varphi_1},\cdots, \frac{1}{\lambda_\mu} \ket{\varphi_\mu} \right] =0 ,
 \label{eq:decay-criterion}
\end{equation}
where we have used Eqs. \eqref{eq:Phi-decomposed}
and \eqref{eq:half_kets}.
The surface state energies $\varepsilon_{\rm sur}$ can be obtained using the above relation which serves as the decay condition for the surface states.

{{The determinant conditions given by Eq. \eqref{eq:decay-criterion-old} or equivalently \eqref{eq:decay-criterion}, can be interpreted in the following way. The generalized eigenvalue problem for the transfer matrix yields the full set of solutions of the system in the absence of boundaries, with no restriction on the energy spectrum. Introducing the surface boundary condition reduces this space by enforcing compatibility between the independent solutions and the boundary. In this process, the solutions cease to be linearly independent, and the resulting constraint is precisely what gives rise to the determinant condition. Under boundary conditions other than the hard surface termination, one can still repeat the procedure starting from Eq. \eqref{eq:surface_state_Psi}, which would again lead to a determinant condition equivalent to the imposed boundary constraint, though possibly with a different explicit form.
}}

Before proceeding to the results section, it is important to note that our method can be generalized to include inter-layer hopping terms beyond nearest neighbors. This can be achieved by reformulating the model to combine two adjacent layers into a single "superlayer," with each superlayer consisting of two layers. By doing so, we can effectively account for next-nearest neighbor hopping terms, reducing the problem to one of nearest-neighbor hopping between these superlayers. Although this method can be further extended to incorporate models with even longer-range hoppings, such an extension would increase the dimensionality of the matrices involved, potentially adding to the computational complexity.

\section{Results}

This section presents the exact results for the surface states of two important lattice models used to study topological insulating and metallic phases: (a) the BHZ model and (b) the two-band model for Weyl semimetals, derived using our transfer matrix formulation.
We also consider the slab geometry composed of multiple layers for both models and numerically compute the dispersion relations using exact diagonalization of their tight-binding Hamiltonians. The transfer matrix results show complete agreement with the numerical results for sufficiently wide slabs, not only for the surface states but also for the projection of bulk bands across the entire 2D Brillouin zone of the surface.

\subsection{3D Bernevig-Hughes-Zhang model}

\begin{figure*}[t!]
\includegraphics[width=0.99\textwidth]{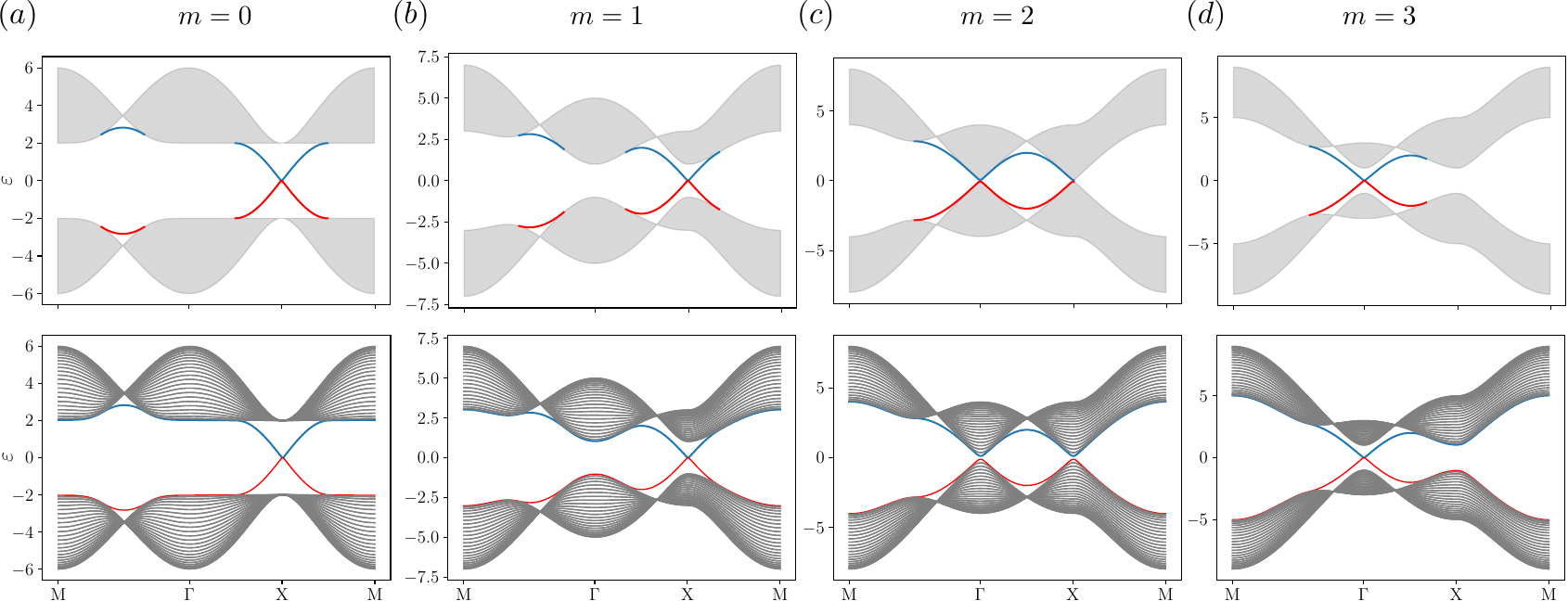}
\caption{
Band structure of surface states and bulk modes
of the 3D BHZ model. The energies are shown along the path between high symmetry points ${\rm \Gamma}$, M, and X in the surface-projected 2D Brillouin zone. The upper panels show the analytical results obtained from the transfer matrix technique, and the lower panels are corresponding results calculated from the direct numerical solution of a tight-binding model in a slab geometry with 50 layers parallel to the surface. The bulk bands are in grey and the surface states are shown in red and blue.
{{(a) and (b) show the weak topological phase, with low-energy surface states present only at the ${\rm X}$ point. (c) marks the gap-closing critical point that signals the transition from the weak to the strong topological phase (STP). Finally, (d) illustrates the resulting STP phase, featuring surface modes at the ${\rm \Gamma}$ point.}}
}
\label{fig_BHZ}
\end{figure*}

We first consider the 3D BHZ model given by the lattice-regularized Hamiltonian \cite{zhang2009topological,Zhang2010model,Budich2023},
\begin{equation}
{\cal H}_{\mathrm{BHZ}} (\mathbf{k})= {M}_{\bf k}
\Gamma_0 +2\sum_{i=1}^3 \lambda_i\sin k_i \Gamma_i 
\end{equation}
with Gamma matrices
$\Gamma_0 = \sigma_0 \otimes \tau_z$,
$\Gamma_1 = \sigma_z \otimes \tau_x$,
$\Gamma_2 = \sigma_0 \otimes \tau_y$,
$\Gamma_3 = \sigma_x \otimes \tau_x$ (all terms are mutually anti-commuting). The diagonal terms are given by energy expression
${M}_{\bf k}=m-2\sum_{i}t_i \cos k_i$.
Using the generic expression for the bulk states of the multi-band model given in App. \ref{App:Bulk},
the bulk state energies of the 3D BHZ model read 
\begin{align}
    \varepsilon^{\rm bulk}_{\bf k} = \pm \sqrt{{M}_{\bf k}^2+4\sum_{i=1}^3\lambda_i^2 \sin^2 k_i}.
    \label{eq:bulk_energy_BHZ}
\end{align}
This Hamiltonian can show weak and strong topological phases as well as a trivial insulating phase \cite{Budich2023}. Assuming all equal hopping terms $t_i=t$,
the strong topological phase (STP) which is robust against disorder 
appears for $2<|m/t|<6$ with surface states at the vicinity of ${\rm\Gamma}$ point. The weak topological phase (WTP) exist for $|m/t|<2$ with surface modes at the vicinity of ${\rm X}$ point. The system undergoes gap closing and topological phase transitions at $|m/t|=2,6$ and we have a trivial phase when $|m/t| >6$.

Considering a slab geometry with surfaces parallel to $z$ direction,
the transfer matrix can be constructed by re-writing the expressions for the momentum-dependent coefficients of the Hamiltonian
by treating the terms involving momentum $k_y$ as the translation operators ${\rm T}_{\pm y}$. This then leads to the recursive Eq. \eqref{eq:transer1}
with
\begin{align}
    \mathcal{M} &= m_{\bf k}\Gamma_0+2\sum_{i=1}^2\lambda_i\sin k_i , \\
    \mathcal{T} &= -\left( t_z \Gamma_0 + i \lambda_z \Gamma_3\right),
\end{align}
in which $m_{\bf k} = M_{\bf k}+2 t_z \cos k_z = m-2\sum_{i=1}^2t_i \cos k_i$.

Following the recipe of previous section for transfer matrix and as detailed in the  App. \ref{app:3d-BHZ}, we finally find the following relation
\begin{equation}
    -4 \nu  {m_{\bf k}} \pm \theta  \left({m_{\bf k}}^2-4 (\nu+1)\right) + 2 \theta ^2 {m_{\bf k}}\pm \theta ^3 = 0 ,
    \label{eq:decay-criterion-BHZ}
\end{equation}
for the energies of surface states and projected bulk states of the 3D BHZ Hamiltonian. Here, the energy dependency comes through the term $\theta = \sqrt{\varepsilon^2-|\Lambda_{\bf k}|^2}$
with $\Lambda_{\bf k}=2(\lambda_x \sin k_x+i \lambda_y \sin k_y)$.
The other two parameters are given by
$m_{\bf k}=m-2\sum_{i=1,2} t_i\cos k_i$
and $\nu = (t_z^2- \lambda_z^2)/t_z$. Note that we have considered the termination in the $z$-direction
such that the surface states are localized at the $z=0$ surface, for instance.
Eq. \eqref{eq:decay-criterion-BHZ} which is a cubic equation for $\theta$ 
can be solved to obtain the surface state energies.
But the full expression of the solutions in generic cases are quite lengthy 
and therefore, they are only presented in the App. \ref{app:3d-BHZ}.

Considering a special case of $\lambda_z=\pm t_z$ which implies $\nu=0$, 
the solutions for $\theta$ are $\theta=0$, and $\theta=\pm (m_{\bf k}\pm2)$.
Substituting these results into the expression of transfer matrix eigenvalues $\lambda$, 
we find that only the solutions with $\theta=0$ are proper surface states with decaying 
nature for which $\lambda$ is $m_{\bf k}$ or $1/m_{\bf k}$ 
depending on whether $m_{\bf k}<1$ or $m_{\bf k}>1$, respectively.
Therefore, surface state energies for the 3D BHZ model reads, 
\begin{equation}
    \varepsilon^{\rm sur}(k_x,k_y) = \pm \Lambda_{\bf k} =
    \pm 2\sqrt{\lambda_x^2 \sin^2 k_x+\lambda_y^2 \sin^2 k_y}.
\end{equation} 
We can also check that the bulk states maximum and minimum energies, 
when projected onto the 2D Brillouin zone of the surface, are given by
\begin{align}
\varepsilon^{\rm bulk}(k_x,k_y) &= \pm \sqrt{\Lambda_{\bf k}^2+(m_{\bf k}\pm2 )^2} \nonumber\\
 &=\pm 2\sqrt{\lambda_x^2 \sin^2 k_x+\lambda_y^2 \sin^2 k_y +(\frac{m_{\bf k}}{2}+ \pm 1 )^2},\label{eq:BHZ-bulk-proj}
\end{align}
where we have all four possible combinations of the two signs which correspond to four energy dispersion relations. 
It is noteworthy that the maximum and minimum energies of bulk states can be determined through two distinct methods. Firstly, they can be obtained by inspecting the energy expression of bulk states, as given in Eq. \eqref{eq:BHZ-bulk-proj}. Secondly, and more intriguingly, the energy band edges can also be derived from alternative solutions for $\theta$, wherein the transfer matrix eigenvalue reaches unit amplitude ($|\lambda|=1$). This indicates that these solutions exhibit non-decaying characteristics, effectively representing bulk states projected onto the 2D Brillouin zone of the surface.

We now compare analytical expressions for the surface states and bulk band edges projected onto the surface Brillouin zone with those obtained from direct numerical diagonalization as presented in Fig. \ref{fig_BHZ}. For the numerical analysis, we considered the tight-binding Hamiltonians for a slab geometry consisting of 50 layers. The hopping parameters were set to $t_i = \lambda_i = 1$, and the mass parameter $m$ was varied to explore different phases of the model (WTP, STP, and trivial insulator). As illustrated in the figure, the analytical expressions perfectly match the numerical results. Panels (a) and (b) display the WTP with surface states present at the vicinity ${\rm X}$ point. Panel (c) represents the gap closing point, marking the transition from the WTP to the STP, which is further shown in panel (d).
We also perform an error analysis, detailed in Table \ref{table}, to confirm the high level of agreement between the exact and numerical approaches. We use the root mean squared error (RMSE) to quantify the deviations, defined as:
\begin{align}
\text{RMSE} = \sqrt{\frac{1}{\mathcal{N}} \sum_{i=1}^{\mathcal{N}} (\varepsilon^{\text{num}}_{{\bf k}_i} - \varepsilon^{\text{exact}}_{{\bf k}_i})^2 }.
\end{align}
The RMSE values are small, indicating precise alignment between the numerical results and the exact analytical solutions. As expected, the deviations decrease with an increasing number of layers.

\begin{table}[t]
\centering
\begin{tabular}{|c|c|c|c|c|c|}
\hline
\textbf{} & {\# layers} & \textbf{$m=0$} & \textbf{$m=1$} & \textbf{$m=2$} & \textbf{$m=3$} \\ \hline
\hline
\multirow{3}{*}{\rotatebox{90}{Bulk}} & 10 & $4.2\times 10^{-2}$ & $4.3\times 10^{-2}$ & $4.5\times 10^{-2}$ & $4.8\times 10^{-2}$ \\ \cline{2-6} 
 & 20 & $1.1\times 10^{-2}$ & $1.2\times 10^{-2}$ & $1.2\times 10^{-2}$ & $1.3\times 10^{-2}$ \\ \cline{2-6} 
 & 40 & $2.9\times 10^{-3}$ & $3.0\times 10^{-3}$ & $3.1\times 10^{-3}$ & $3.3\times 10^{-3}$
 \\ \hline
 \hline
\multirow{3}{*}{\rotatebox{90}{Surface}} & 10 & $3.9\times 10^{-3}$ & $4.8\times 10^{-3}$ & $6.2\times 10^{-2}$ & $3.4\times 10^{-3}$ \\ \cline{2-6} 
 & 20 & $7.6\times 10^{-4}$ & $9.2\times 10^{-4}$ & $2.4\times 10^{-2}$ & $6.7\times 10^{-4}$ \\ \cline{2-6} 
 & 40 & $1.4\times 10^{-4}$ & $1.7\times 10^{-4}$ & $9.5\times 10^{-3}$ & $1.3\times 10^{-4}$ \\ \hline
\end{tabular}
\caption{RMSE between numerical results for slabs of varying widths (number of layers) and exact analytical results for the 3D BHZ model. The table presents the results separately for bulk band edges and surface modes. Overall, the deviations are small and decrease as the number of layers increases.}
\label{table}
\end{table}

\begin{figure*}[t!]
\includegraphics[width=0.99\textwidth]{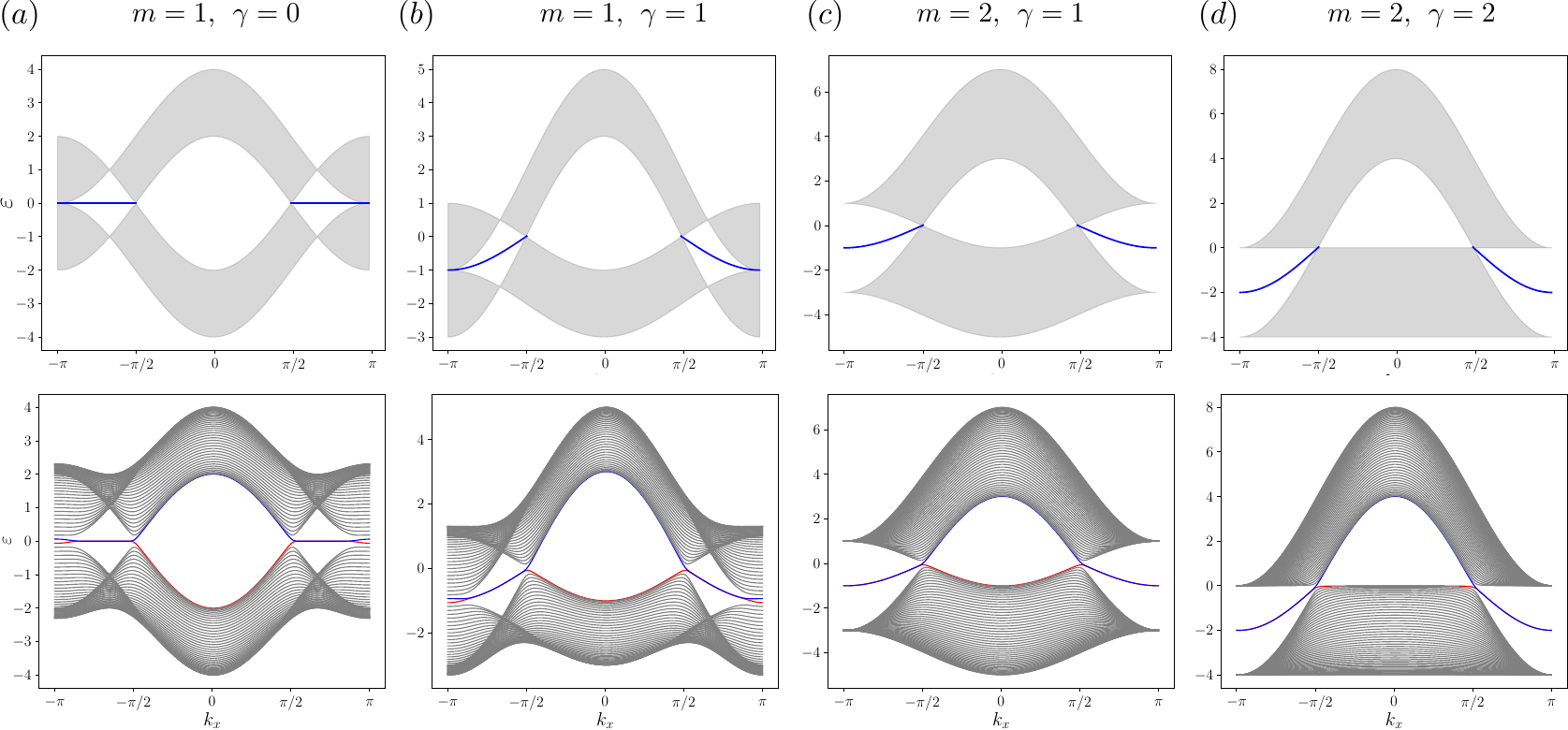}
\caption{
Band structure of surface states and bulk modes
of the 3D simple Weyl model consisting of only two bands. 
Similar to Fig. \ref{fig_BHZ}, the upper panels show the analytical results obtained from the transfer matrix technique, and the lower panels show the corresponding results calculated from the direct numerical solution. The bulk bands are in grey and Fermi arc states are shown in blue and red. {{In the transfer matrix results (upper panels), the two degenerate Fermi arc surface states overlap exactly, so only a single blue line is visible. In contrast, in the numerical results for finite slabs (lower panels), the two branches deviate slightly near the Brillouin zone edges and where they merge with the bulk states, leading to the appearance of small gaps due to finite-size effects. Panels (a–c) correspond to a type-I Weyl semimetal with different tilts, controlled by the parameter $\gamma$, and with the band-edge gap tuned by the parameter $m$. Panel (d) corresponds to the critical value $\gamma=2$, beyond which the system enters the type-II Weyl semimetal phase. We set the two hopping parameters to $t = t_x = 1$ in all panels and used a slab geometry with 50 layers for the numerical calculations shown in the lower panels.}}  
}
\label{fig:simple-weyl}
\end{figure*}

\subsection{Weyl semimetal model}
The minimal lattice model for a WSM consisting of two bands and only two Weyl nodes is given by \cite{McCormick2017}
\begin{equation}
    \hat{\mathcal{H}}_{\text{WSM-I}} (\mathbf{k}) = \sum_{i=0}^3 d_i(\mathbf{k})\hat{\sigma}_i ,
    \label{simple_weyl}
\end{equation} where
\begin{equation}
    \begin{matrix}
        d_0&=& \gamma[\cos(k_x)-\cos(k_0)], \\
         d_1 &=& m[\cos(k_y)+\cos(k_z)-2]\\
        &+& 2t_x[\cos(k_0)-\cos(k_x)], \\ 
        d_2&=& -2t\sin(k_y), \\ 
        d_3  &=& -2t\sin(k_z).
    \end{matrix} 
\end{equation} 
The bulk band energies dispersion for this model is
\begin{equation}
    \varepsilon^{\rm bulk}_{\bf k} = d_0 \pm \sqrt{\sum_{i=1}^3 d_i^2}
\end{equation}
{{
The Weyl nodes of the model are located at ${\bf K}_{\pm}=(\pm k_{0},0,0)$, where the vector ${\bf d}({\bf k})$ vanishes identically, independent of the parameters $m$, $t$, and $t_{x}$. In the vicinity of these nodes, the Hamiltonian can be expanded to linear order in the deviation ${\bf q}={\bf k}-{\bf K}_{\pm}$, yielding
\begin{equation}
\hat{\mathcal{H}}_{\rm low}({\bf q}) = \mp \sin k_{0}\: q_{x} (\gamma \, \hat{\sigma}_{0} - 2t_{x}\, \hat{\sigma}_{z}) 
- 2t (q_{y} \hat{\sigma}_{y} + q_{z} \hat{\sigma}_{z}).
\end{equation}
This simplified low-energy model has the additional advantage of capturing the transition between type-I and type-II Weyl semimetals. In the Hamiltonian \eqref{simple_weyl}, the parameter $\gamma$ controls the tilt of the Weyl cone, with the critical value $\gamma=2t$ marking the boundary between the type-I phase ($\gamma < 2t$) and the type-II phase ($\gamma > 2t$).
}}

Assuming a finite slab geometry with surfaces parallel to $y$ direction, and introducing translation operator similar to the previous example, the 
intra- and inter-layer terms of the recursive relation for the wavefunction components are given by
\(\mathcal{T}=\frac{m}{2}\hat{\sigma}_1+it\hat{\sigma}_2\)
and \(\mathcal{M}=d_0(\mathbf{k})\hat{\sigma}_0+h_1(\mathbf{k})\hat{\sigma}_1+d_3(\mathbf{k})\hat{\sigma}_3\) where 
\begin{align}
h_1&= d_1  -m\cos(k_y) \nonumber\\ 
&= -m[2-\cos(k_z)]-2t_x[\cos(k_x)-\cos(k_0)].
\end{align}
Applying the formulation of Sec. \ref{sec:formulation} as detailed in App. \ref{app:simple_weyl}, the dispersion of Fermi arc states and projection of bulk states to the surface Brillouin zone are given by
\begin{align}
    &\varepsilon^{\rm sur}(k_x,k_z) = d_0 \pm |d_3|, \\
    &\varepsilon^{\rm bulk}(k_x,k_z) = d_0 \pm \sqrt{d_3^2 + (h_1 \pm m)^2},
\end{align}
respectively.

The results derived from the analytical equations above, along with those obtained from direct numerical diagonalization of the tight-binding Hamiltonians for a slab geometry consisting of 50 layers, are presented in Fig. \ref{fig:simple-weyl}. The figure illustrates consistency between the exact expressions and numerical results for the surface modes and the bulk band edges (by projection to the 2D Brillouin zone of the surface).

\begin{figure}[t!]
\includegraphics[width=0.9\linewidth]{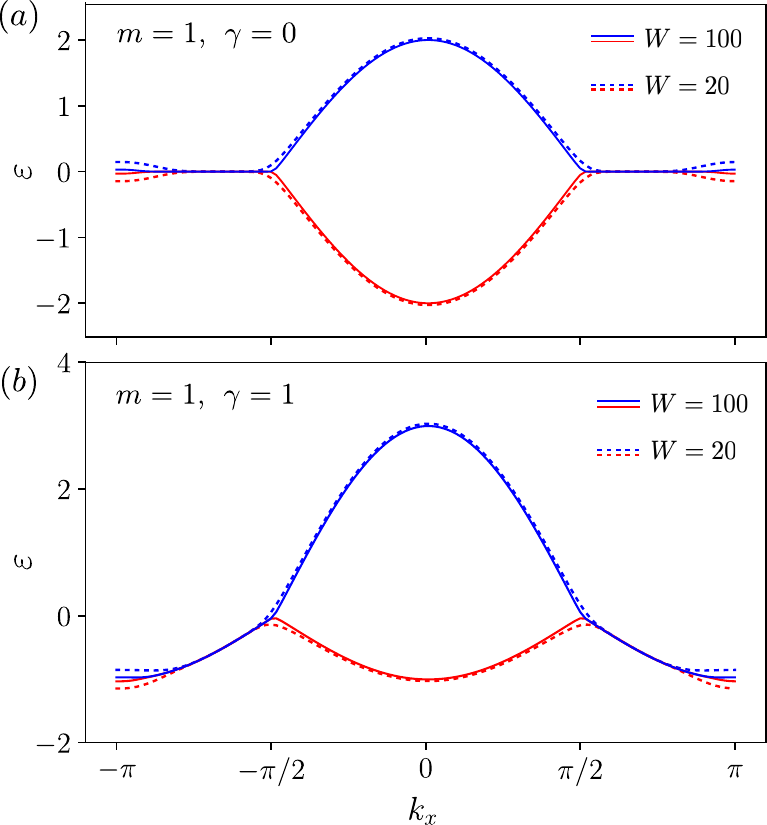}
\caption{{
Dispersion of Fermi arc states and their continuation into the bulk states for $k_x \in (-\pi/2,\pi/2)$, obtained numerically for slab geometries with two different widths. As shown, increasing the slab width reduces the small finite-size gaps between the two branches of the Fermi arc states. For sufficiently wide slabs, these gaps vanish, and the results converge to the exact solution provided by the transfer matrix method. Panels (a) and (b) correspond to different tilt parameters, $\gamma=0$ and $\gamma=1$, respectively.
}}
\label{fig_new}
\end{figure}

{{As seen in the lower panels (a) and (b) of Fig.~\ref{fig:simple-weyl}, corresponding to the direct numerical diagonalization of finite slabs, the Fermi arc states exhibit small gaps near the Brillouin zone edges due to finite-size effects. These gaps originate from the limited slab width and systematically decrease as the width increases. In the limit of an infinitely wide slab, the numerical results converge to those obtained from the transfer matrix method, which provides the exact solution. This behavior is further illustrated in Fig.~\ref{fig_new}, where we compare slabs of widths $W=20$ and $W=100$ for different tilt parameters. As the figure shows, the finite-size gaps shrink with increasing slab width, confirming that they are not intrinsic features but rather artifacts of the finite geometry.}}

\section{Discussion}
We have seen how the exact formalism based on transfer matrix techniques 
can be used to derive topological surface states and Fermi arc states in 3D tight-binding lattice models that exhibit topological insulating and semimetallic behaviors. The main ingredients of this method is to combine generalized transfer matrix technique that neither relies nor
requires non-invertibility of inter-layer hopping matrices, and a simple form of boundary condition based on decaying criteria for surface modes in a semi-infinite geometry.
This method, in fact, is not limited to exploring lattice models exhibiting topological phases, and can be equally used for obtaining non-topological surface states. 

{
We emphasize that the issue of non-invertible interlayer hopping matrices only arises in earlier formulations of the transfer matrix, where explicit matrix inversion was required. Our construction in Eqs. \eqref{eq:S_matrix} and \eqref{eq:S_matrix_def} avoids such inversion altogether, thereby eliminating the associated numerical instabilities. Degeneracies of the interlayer hopping matrix  ${\mathcal T}$ are not problematic, as it appears only in the off-diagonal blocks. However, degeneracies of the full transfer matrix may occur depending on the specific forms of ${\mathcal T}$ and ${\mathcal M}$. In those cases, numerical stability can be maintained using standard techniques such as Gram–Schmidt orthogonalization in the eigenvalue calculations.}

One important point is that the success of our method in providing exact expressions for surface states using the transfer matrix technique, depends on the exact solvability of the eigenvalue equations and the boundary conditions. In particular, increasing the number of bands leads to larger matrices and higher-degree secular equations. Since polynomial equations of degree five or higher do not generally have exact solutions, there is no guarantee that exact surface dispersion relations can always be found. In these situations, a numerical framework based on the transfer matrix can be used instead of numerically diagonalizing real-space Hamiltonians for finite geometries in one direction. 

{{
\subsection{Computational scaling of the method}}}

The computational complexity of the transfer matrix method employed here is determined by the size of the matrix $\mathbb{T}$, which is twice the number of bands, $n_{\text{bands}}$. Consequently, when utilizing a basic diagonalization algorithm, the time complexity of the calculations is $\mathcal{O}(n_{\text{bands}}^3)$. Notably, this complexity is \emph{entirely independent} of the number of layers in a finite slab configuration. In contrast, direct diagonalization for a slab geometry with $N$ layers, even when employing efficient iterative methods such as Lanczos or Arnoldi, scales as $\mathcal{O}(k N n_{\text{bands}})$, where $k$ represents the number of eigenvalues of interest (in our case, those corresponding to surface modes). Overall, if we focus on the scaling with $N$ and assume that $n_{\text{bands}}$ remains small (as is typical for most lattice models of interest), the transfer matrix method achieves an $\mathcal{O}(1)$ computational cost. In contrast, the best-case efficiency for direct diagonalization of the slab Hamiltonian is $\mathcal{O}(N)$.

Moreover, diagonalization of the real-space Hamiltonian is limited to finite-sized slabs and requires size-scaling analysis to extrapolate results for surface states in semi-infinite or very wide geometries. In contrast, the transfer matrix method allows for exact analytical or numerical analysis of surface physics, which surpasses brute-force approaches, particularly when dealing with semi-infinite geometries or isolated surface states. The development of a numerical method based on the transfer matrix is left for a future work.

\vspace{0.5cm}
{{\subsection{Comparison with recursive green’s function}}}

{{
Both the recursive Green’s function approach and the transfer-matrix formalism are widely used to study the surface physics of semi-infinite systems, but they differ in their mathematical formulation and in the type of information they provide. Recursive Green’s function methods, such as the Lopez Sancho algorithm \cite{sancho1984quick,sancho1985highly}, work by recursively integrating out bulk layers and renormalizing their effect into the interface layer. This procedure yields the surface Green’s function, from which spectral properties and local densities of states can be extracted. However, since the bulk layers are effectively absorbed into the surface self-energy, information about the detailed penetration of surface states into the bulk is not directly retained.

By contrast, in the transfer-matrix method the relation between wavefunction amplitudes of adjacent layers is preserved through the eigenvalues of the transfer matrix. This provides detailed information about the decay and penetration depth of surface states perpendicular to the interface, in addition to their dispersions. From a computational standpoint, both methods avoid the need to build and diagonalize very large slab Hamiltonians and thus achieve efficiency that is independent of slab thickness. The recursive Green’s function method, however, is iterative in nature and requires convergence, whereas the transfer-matrix method reduces the problem to solving a generalized eigenvalue equation of fixed size, which is non-iterative and direct.
\par
Finally, recursive Green’s function methods are well established and widely implemented, with the advantage of naturally incorporating disorder and interactions via self-energies. The transfer-matrix method, in contrast, is less developed in this regard but provides more direct analytic access to the detailed structure of surface states beyond the first interface layer.
}}

\vspace{0.5cm}
{{\subsection{Applications to interfaces and heterostructures}}}

Now, our results can be employed to investigate surface physics and interfaces involving topological phases, providing valuable insights into the surface physics of various topological states beyond the low-energy limit.
Specifically, the analytical results for the surface states dispersion throughout the surface Brillouin zone can be applied to explore engineered topological phases at the interfaces 
and heterostructures. These interfaces, where different topological phases or a topological insulator and a conventional material interact, can give rise to novel quantum states \cite{bernevig2006quantum,konig2007quantum,Hasegawa2011,Burkov2011Weyl,Qi2013,katmis2016ferromagnetic,Weber_2024,Mertig2013,kotetes2013classification}, and hold promise for electronics and spintronics application \cite{chang2013experimental,bernevig2022progress,mellnik2014spin,Moghaddam2020}. Such engineered phases are particularly significant in the context of topological superconductivity, where the proximity effect between a superconductor and a topological insulator can induce a superconducting phase that inherits the topological properties 
\cite{FuKane2008Majorana,
MajoranaWire_Oreg,MajoranaWire_Sau,
Alicea2010,Stanescu2011, mourik2012signatures,nadj2014observation,
sato2017topological,frolov2020topological,flensberg2021engineered}.

As a concrete example, one can construct an effective model of the interface based on the exact surface dispersions derived here and introduce the effective superconducting pairing between them. This elevates the model beyond typical studies that use low-energy surface-mode dispersions, often approximated by an effective Dirac Hamiltonian. 
{{In practice, for heterostructures under the assumption of weak coupling, one can first treat the two sides of the interface as independent surfaces, obtain effective models for each using the transfer-matrix method, and then introduce the inter-surface coupling to capture interactions between the layers. This approach allows the study of interface physics, for example between materials with different ordering such as magnetism and superconductivity.}} Furthermore, it bridges the gap between effective low-energy models and fully numerical models based on tight-binding or ab initio methods, enabling a more comprehensive exploration of interface-induced topological superconductivity. Similar strategies can be applied to other engineered topological phases at the boundaries between a topological material and a non-topological system. A notable example is interfacial magnetic topological insulators formed by the proximity of time-reversal-invariant 3D topological insulators and materials with magnetic ordering.

{{Our formalism provides not only the surface-state dispersions but also the corresponding eigenstates, allowing direct computation of physical quantities such as the contributions of surface states to conductance or their spin textures. This feature highlights the broader applicability of our approach and clarifies the connection between the calculated surface states and measurable physical properties.}}

\vspace{0.5cm}
{
{\subsection{Extensions and Outlook}
}}

{
{So far we have seen that The formalism can be applied directly to a generic non-interacting lattice models including multi-orbital cases at least in a numerical manner. Extensions to interacting systems could be pursued within mean-field approximations, with more advanced many-body schemes left for future work. The method is most naturally suited to semi-infinite geometries, but in principle could be adapted to finite slabs by incorporating two-sided boundary conditions. Effects of surface roughness or disorder may also be included at an averaged level, for instance through disorder-induced self-energies. In addition, the transfer-matrix construction could be combined with DFT–Wannier-based tight-binding models \cite{Wannier-rmp2012}, enabling first-principles studies of surface states and their comparison with experimental data. These extensions, while beyond the scope of the present paper, represent promising directions for broadening the applicability of the transfer-matrix approach to more realistic systems.
\par
%
Beyond these theoretical considerations, we note that key features obtained from our approach, such as the merging of surface-state dispersions into the bulk continuum at higher energies, are directly observable in ARPES experiments, where similar behavior is routinely reported in topological semimetals and related systems. ARPES measurements have revealed these characteristic features in a variety of topological insulators, including Bi$_2$Se$_3$ and related compounds \cite{Bi2Se3_PRL,xia2009observation,Hofmann_2021,Dil_2019}, and later in magnetic topological insulators such as MnBi$_2$Te$_4$ \cite{Otrokov2019}. Likewise, for Weyl semimetals, ARPES results have been reported for a range of materials, including TaAs, NdAlSi, and CeAlSi \cite{jia2016weyl,li2023emergence,cheng2024tunable}. Although the models studied here are intentionally minimal, they already reproduce characteristic signatures of ARPES spectra, supporting the view that the formalism captures robust aspects of surface electronic structure. As mentioned above, an important future step will be to combine the method with material-specific model Hamiltonians to enable quantitative comparison with experimental data.
}}

In summary, we have applied an exact formalism based on transfer matrix techniques to derive topological surface states and Fermi arc states in 3D tight-binding lattice models that exhibit topological insulating and semimetallic behaviors. This formulation allows us to decompose both surface and bulk states in terms of the eigenstates of this generalized eigenvalue problem, serving as a substitute for the traditional transfer matrix method. By applying boundary conditions for surface termination, we derive analytical expressions for surface state energies and the edges of bulk bands in the projected 2D Brillouin zone. Our method has been successfully applied to well-known models such as the 3D Bernevig-Hughes-Zhang model and a widely studied two-band model for Weyl semimetals. These analytical results provide a more efficient alternative to fully numerical methods, particularly in cases requiring surface state properties beyond the low-energy regimes.

\appendix
\section{Bulk states of the generic multi-band model}
\label{App:Bulk}

We consider a set of mutually anti-commuting gamma 
matrices $\Gamma_i$ which square to identity which is divided 
into two sets $\{\Gamma_i^{(1)}\}$, $\{\Gamma_j^{(2)}\}$, respectively.
The anti-commutation relations are given by
\begin{align}
    \{\Gamma_i^{(q)},\Gamma_j^{(q')} \} = 2 \delta_{ij}\delta_{qq'} \, \mathbbm{1},       
\end{align}
So far the division of the gamma matrices into two sets seems arbitrary. 
Now we assume there are two extra matrices 
$\Xi^{(1)}$ and $\Xi^{(2)}$ which 
\emph{commute} with each other ($\left[ \Xi^{(1)},\Xi^{(2)}\right]=0$) and square to identity.
Additionally, we have commutation and anti-commutation relations
between $\Xi^{(q)}$ and $\Gamma_i^{(q')}$ as follows:
\begin{align}
    &[\Xi^{(q)},\Gamma_i^{(q)} ] = 0, \\
    &\{\Xi^{(q)},\Gamma_i^{(q')} \} = 0, \qquad (q\neq q')
\end{align}
Then, we can write the generic $k$-space Hamiltonian expressed in terms of all these gamma matrices 
and an additional identity term as
\begin{align}
   {\cal H} = \delta\mathbbm{1}+ \sum_{q=1,2} \left( {\bf h}_q\cdot{\bm\Gamma}^{(q)}
    + \xi_q \Xi^{(q)} \right),
\end{align}
where coefficients ${\bf h}_q = h_{q,i}\hat{\bf e}_i$ and $\xi^{(q)}$ 
depend on the momentum ${\bf k}$.
Now, as we will see below, the energy spectrum of this Hamiltonian can be 
exactly obtained in terms of the coefficients $h_i^{(q)}$ and $\xi^{(q)}$.
To this end, we first take the square of the Hamiltonian after subtracting 
the identity term $\delta\mathbf{I}$ as
\begin{align}
    ({\cal H}-\delta\mathbbm{1})^2 &= 
    \left(\sum_{q=1,2} {\bf h}_q \cdot{\bm\Gamma}^{(q)}   
    \right)^2
    +\left(\sum_{q=1,2} \xi_q \Xi^{(q)} \right)^2
     \nonumber \\
    &+ 
    \sum_{q,q'=1,2}
    \left\{
         {\bf h}_q \cdot{\bm\Gamma}^{(q)},
         \xi_{q'} \Xi^{(q')} 
    \right\}\\
    &=
    \sum_{q=1,2} \left( |{\bf h}_q|^2+ |\xi_q |^2 \right) \mathbbm{1}+ 
    2\xi_1 \xi_2 \,\Xi^{(1)}\Xi^{(2)} \nonumber \\
    &+
    2\sum_{q=1,2} \xi_q \Xi^{(q)}
    {\bf h}_q \cdot{\bm\Gamma}^{(q)},
\end{align}
where to obtain the last line we have used the anti-commutation and commutation relations between different matrices.
Let's bring the first term of the last line, which is proportional to identity, 
to the left-hand side and square it again. This results in
\begin{align}
    &\frac{1}{4}\left[ ({\cal H}-\delta\mathbbm{1})^2 - 
    \sum_{q=1,2} \left( |{\bf h}_q|^2+ |\xi_q |^2 \right) \mathbbm{1} \right]^2
    \nonumber \\
    &=
    \left(\sum_{q=1,2} \xi_q \Xi^{(q)}
    {\bf h}_q \cdot{\bm\Gamma}^{(q)}\right)^2
    +\xi_1^2 \xi_2^2 \, \mathbbm{1} \nonumber \\
    &+
    \left\{    
    \sum_{q=1,2} \xi_q \Xi^{(q)}
    {\bf h}_q \cdot{\bm\Gamma}^{(q)}
    ,
    \xi_1 \xi_2 \,\Xi^{(1)}\Xi^{(2)}
    \right\}, 
    \label{eq:app1_fourth_power}
\end{align}
where we have used the fact that $\left(\Xi^{(1)}\Xi^{(2)}\right)^2 = \mathbbm{1}$.
As we can easily verify, the new sets of matrices given by $\Xi^{(q)}\Gamma^{(q)}_i$
are also mutually anti-commuting and square to identity as
\begin{align}
    \left\{ \Xi^{(q)}\Gamma^{(q)}_i, \Xi^{(q')}\Gamma^{(q')}_j \right\} = 
    2\delta_{ij}\delta_{qq'}\mathbbm{1}.
\end{align}
Therefore, the first term on the right-hand side of 
Eq. \eqref{eq:app1_fourth_power} simplifies to
\begin{align}
    \left(\sum_{q=1,2} \xi_q \Xi^{(q)}
    {\bf h}_q \cdot{\bm\Gamma}^{(q)}\right)^2
    =
    \sum_{q=1,2} \xi_q^2 |{\bf h}_q|^2 \: \mathbbm{1} .
    \end{align}
Furthermore, as one can verify by direct inspection, 
the last term in the right-hand side of Eq. \eqref{eq:app1_fourth_power}
vanishes identically. Therefore, we have
\begin{align}
    &\frac{1}{4}\left[ ({\cal H}-\delta\mathbbm{1})^2 - 
    \sum_{q=1,2} \left( |{\bf h}_q|^2+ |\xi_q |^2 \right) \mathbbm{1} \right]^2
    \nonumber \\
    &=
    \left(
    \sum_{q=1,2} \xi_q^2 |{\bf h}_q|^2
    +\xi_1^2 \xi_2^2 
    \right)
    \: \mathbbm{1}.
\end{align}

Since all the matrices apart from the Hamiltonian are now 
proportional to identity, we can deduce the characteristic equation for 
the eigenvalues of the Hamiltonian as
\cite{note1},
\begin{align}
    \left(\varepsilon - \delta\right)^2
    &-\sum_{q} \left( \xi_q^2 +|{\bf h}_q|^2 \right) 
    = \nonumber\\
    &\pm 2\sqrt{
        \sum_{q} \xi_q^2 |{\bf h}_q|^2
        + \xi_1^2 \xi_2^2
    },
\end{align}
and therefore the energies are given by
\begin{align}
    &\varepsilon_{\zeta \zeta' } = \delta 
    +\zeta \nonumber\\
    &\:
    \times \sqrt{
        \sum_{q} \left(\xi_q^2 +|{\bf h}_q|^2\right)
         +
         2\zeta'\sqrt{
            \sum_{q} \xi_q^2 |{\bf h}_q|^2
            + \xi_1^2 \xi_2^2
        }
    }.
\end{align}

\section{Transfer matrix for the 3D BHZ model}
\label{app:3d-BHZ}
We first work out the characteristic equation for the transfer matrix eigenvalues by inserting the 
matrices ${\cal M}$ and ${\cal T}$ of 3D BHZ model into Eq. \eqref{eq:main-characteristic} and evaluating it which results in:
\begin{align}
\left[- \frac{\nu}{2} \left( \lambda^4 + 1 \right) 
+ m_{\bf k}\left( \lambda^3 + \lambda \right) 
+  \lambda^2 (\nu - \zeta)\right]^2 
= 0 ,
\end{align}
in which
\begin{align}
  \nu &= \frac{t_z^2 - \lambda_z^2}{ t_z},
\\
  \zeta &= \frac{m_{\bf k}^2+ 4 t_z^2 - \varepsilon^2+|\Lambda_{\bf k}|^2}{2t_z},
  \Lambda_{\bf k}=2\sum_{i=1}^2\lambda_i \sin k_i.
\end{align}
The solutions of the characteristic equation above are all two-fold degenerate and can be written as
\begin{align}
    \begin{array}{cc}
         \lambda_{1,2} = \frac{1}{2}\left(\chi_+\pm\sqrt{\chi_+^2-4}\right)&  \\
         \lambda_{3,4} = \frac{1}{2}\left( \chi_-\pm\sqrt{\chi_-^2-4}\right) & 
    \end{array}, \label{eq:lambdas_BHZ}
\end{align}
where $\chi_\pm$ are the two roots of
\begin{equation}
    - \frac{\nu}{2}\: \chi^2
+ m_{\bf k} \:\chi 
+  (2\nu - \zeta) 
= 0 ,
\end{equation}
For any $\lambda$ we have two different eigenstate whose
lower part using the decomposition in \eqref{eq:Phi-decomposed} are
\begin{align}
    \ket{\varphi}_{1,\lambda} \propto \begin{pmatrix}
    (\lambda^2 - 1) \lambda_z \\
    0 \\
    -i \lambda \Lambda_{\bf k} \\
    it_z (1+\lambda^2)+i\lambda (\varepsilon-m_{\bf k})
    \end{pmatrix},
\end{align}
and
\begin{align}
    \ket{\varphi}_{2,\lambda} \propto \begin{pmatrix}
    0 \\
    (\lambda^2 - 1) \lambda_z \\
     -it_z (1+\lambda^2)+i\lambda (\varepsilon+m_{\bf k}) \\
    -i \lambda \Lambda_{\bf k} 
    \end{pmatrix}.
\end{align}

Now, the decaying surface state criteria (boundary condition) described by Eq. \eqref{eq:decay-criterion}, by plugging the eigenstate $\ket{\varphi}_{i,\lambda}$ and after some algebraic manipulations reduces to
\begin{align}
    \frac{\lambda_1+\lambda_2}{\lambda_1 \lambda_2+1}=
    \frac{m_{\bf k}\pm \sqrt{\varepsilon^2-|\Lambda_{\bf k}|^2}}{2t},
\end{align}
Invoking the expressions for the
transfer matrix eigenvalues and after some algebra,
we find
\begin{equation}
    -4 \nu  {m_{\bf k}} \pm \theta  \left({m_{\bf k}}^2-4 (\nu+1)\right) + 2 \theta ^2 {m_{\bf k}}\pm \theta ^3 = 0 ,
\end{equation}
where
$\theta = \sqrt{\varepsilon^2-|\Lambda_{\bf k}|^2}$.

As the two equations are simply related to each other by $\theta\to -\theta$, we can focus only on the equation with positive signs, and the solutions
for the other equations are simply obtained by changing the signs.
For the particular case of $\nu=0$, the equation can be easily factorized as
\begin{equation}
     \pm \theta [\left(\theta\pm{m_{\bf k}}\right)^2-4]= 0 ,
\end{equation}
with five different solutions $\theta=0$, and $\theta = \pm 2 \pm m_{\bf k}$.

For the generic case ($\nu\neq 0$), we first simplify the equation by absorbing the quadratic term via defining a new variable
\begin{align}
    \phi = \theta + \frac{2}{3}m_{\bf k},
\end{align}
in terms of which the equation becomes
\begin{align}
    \phi^3 
    - \phi  \left(4 (\nu +1)+\frac{m_{\bf k}^2}{3}\right)-\frac{2}{27} m_{\bf k} \left(18 (\nu -2)+m_{\bf k}^2\right)=0.\label{eq:cubic-phi}
\end{align}
This new equation has the form of the \emph{depressed cubic equation}
\begin{align}
    \phi^3+3p \phi-2q=0, \label{eq:depressed}
\end{align}
whose solutions are given by
\begin{align}
   \phi =\left\{ \begin{array}{cc}
         & z_- -  z_+ \qquad\qquad~ \quad  \\
         & e^{2\pi i/3} z_- - e^{-2\pi i/3} z_+ \\
         & e^{-2\pi i/3} z_- - e^{2\pi i/3} z_+  \\
    \end{array}\right., \label{eq:phi_solutions}
\end{align}
where
\begin{align}
    z_{\pm} = (\sqrt{q^2+p^3}\pm q)^{1/3}. \label{eq:z_pm}
\end{align}
The two parameter $q$ and $p$ can be obtained by comparing Eqs. \eqref{eq:cubic-phi} and \eqref{eq:depressed} as
\begin{align}
    p &= -\frac{1}{3}\left(4 (\nu +1)+\frac{m_{\bf k}^2}{3}\right) ,
    \label{eq:p}\\
    q &= \frac{1}{27} m_{\bf k} \left(18 (\nu -2)+m_{\bf k}^2\right).
    \label{eq:q}
 \end{align}
So the combination of expressions given by Eqs. \eqref{eq:phi_solutions}-\eqref{eq:q} determines the solutions for the variable $\phi$, and thereby
the energies.

\section{Transfer matrix for the simple Weyl model}
\label{app:simple_weyl}

The generalized eigenvalue form of the transfer matrix
for simple Weyl model \eqref{simple_weyl} can be written as the following
\begin{align}
 {\mathbbm S}-\lambda  {\mathbbm R}=
\begin{pmatrix}    
 \xi+d_3 & \lambda  m_+ +h_1 & 0 & m_- \\
 \lambda m_- +h_1 &  \xi-d_3 & m_+ & 0 \\
 1 & 0 & -\lambda  & 0 \\
 0 & 1 & 0 & -\lambda  
 \end{pmatrix},
\end{align}
in which the following variables are defined for the sake of compactness:
\begin{align}
\begin{array}{cc}
    &\xi = \varepsilon - d_0(k_x)=\varepsilon- \gamma[\cos(k_x)-\cos(k_0)],\\
    &h_1 =  -m[2-\cos(k_z)]-2t_x[\cos(k_x)-\cos(k_0)],\\
    &d_3 = 2t \sin k_z ,\\
    &m_{\pm} = \frac{m}{2}\pm t
    \end{array}
\end{align}

The transfer matrix eigenvalues are then found as
\begin{align}
    \begin{array}{cc}
         \lambda_{1,2} = \frac{1}{2}\left(\chi _+\pm\sqrt{\chi _+^2-4}\right)&  \\
         \lambda_{3,4} = \frac{1}{2}\left( \chi _-\pm\sqrt{\chi _-^2-4}\right) & 
    \end{array}, \label{eq:lambdas}
\end{align}
where
\begin{align}
   \chi_\pm = 
    \frac{\pm\sqrt{4 t^2 h_1^2-\left(m^2-4 t^2\right) \left(-\xi^2+4 t^2+d_3^2\right)}-m h_1}{m^2/2-2 t^2}. \label{eq:chi-s}
\end{align}
The corresponding wavefunctions can be decomposed according to \eqref{eq:Phi-decomposed} in which the lower parts denoted by $\ket{\varphi}$ are also found as
\begin{align}
    \ket{\varphi}_i \propto \begin{pmatrix}
        \lambda_i^2 m_++m_-+\lambda_i  h_1\\
        -\lambda_i  (d_3+  \xi)
    \end{pmatrix}.
\end{align}
Using the above wavefunctions and applying the boundary condition \eqref{eq:decay-criterion}, we find the criterion 
\begin{equation}
    \frac{(\lambda_i-\lambda_j) 
    (d_3+\xi) 
    (\lambda_i \lambda_j m_+ - m_-)
    }{\lambda_i \lambda_j}=0 \label{eq:compact_criterion}
\end{equation}
from which the energies can be obtained.

We first examine solutions of the form $\lambda_i\lambda_j = m_-/m_+$ for non-equal pairs of $\lambda$'s. Specifically, considering the condition $\lambda_2\lambda_4 = m_-/m_+$ and substituting the expressions for the $\lambda$'s given by \eqref{eq:lambdas}, we find the solution
\begin{equation}
    \xi^2 = d_3^2,
\end{equation}
which leads to the corresponding energies
\begin{equation}
    \varepsilon = \gamma[\cos(k_x) - \cos(k_0)] \pm 2t\sin(k_z).
\end{equation}
By substituting these energies into the expressions for the transfer matrix eigenvalues $\lambda_{2,4}$, we verify that they indeed yield decaying solutions over a certain range of momenta, thus corresponding to Fermi arc surface states. Additionally, we observe that inspecting other pairs of $\lambda$'s either yields the same result or fails to produce real (physical) solutions for the energies.

The remaining possibility, as indicated by \eqref{eq:compact_criterion}, involves equal pairs of $\lambda$'s whose absolute values are both less than or equal to 1. However, since $\chi_1 \neq \chi_2$ for almost any generic set of parameters and momenta, this scenario reduces to either $\chi_\pm$ being equal to $\pm 2$, which corresponds to either $\lambda_{1,2} = \pm 1$ or $\lambda_{3,4} = \pm 1$, respectively. Therefore, these solutions correspond to bulk states. By applying the conditions in the expression for $\chi_{\pm}$ given in Eq. \eqref{eq:chi-s}, we obtain the solutions
\begin{equation}
    \xi^2 = d_3^2 + (h_1 \pm m)^2,
\end{equation}
which leads to the four energy dispersions
\begin{equation}
    \varepsilon = d_0 \pm \sqrt{d_3^2 + (h_1 \pm m)^2},
\end{equation}
corresponding to the bulk band edges.

\bibliography{refs.bib}    

\begin{thebibliography}{63}
\bibitem [{not()}]{note1}%
  \BibitemOpen
  \href@noop {} {}\bibinfo {note} {{Technially speaking we are applying
  Cayley–Hamilton theorem, which states any matrix satisfies its own
  characteristic equation. Therefore, this equation is identical to the
  characterstic equation ${\rm det}(\lambda {\mathbbm 1}-{\cal H})=0.$ whereby,
  after expanding the determinant, we have replaced the scalar $\lambda$ with
  the Hamiltonian matrix.}}\BibitemShut {Stop}%
\end{thebibliography}
\end{document}